\documentclass[12pt,preprint]{aastex}

\shorttitle{H. Watanabe et al.}
\shortauthors{H. Watanabe et al.}

\begin{document}

\title{Spectro-Polarimetric Observation of an Emerging Flux Region: Triggering Mechanism of Ellerman Bombs}

\author{H. Watanabe\altaffilmark{1,2}}
\affil{Department of Astronomy, Kyoto University, Sakyo-ku, Kyoto 606-8502, JAPAN}
\email{watanabe@kwasan.kyoto-u.ac.jp}

\author{
R. Kitai\altaffilmark{2},
K. Okamoto\altaffilmark{1,2},
K. Nishida\altaffilmark{2},
J. Kiyohara\altaffilmark{2},
S. UeNo\altaffilmark{2},
M. Hagino\altaffilmark{2},
T. T. Ishii\altaffilmark{2},
and K. Shibata\altaffilmark{2}
} 
\affil{Kwasan and Hida Observatories, Kyoto University, Yamashina-ku, Kyoto 607-8417, JAPAN}

\altaffiltext{1}{ Department of Astronomy, Kyoto University, Sakyo-ku, Kyoto 606-8502, JAPAN }
\altaffiltext{2}{ Kwasan and Hida Observatories, Kyoto University, Yamashina-ku, Kyoto 607-8417, JAPAN }


\begin{abstract}

High spatial resolution observation of an emerging flux region (EFR) was done using a vector magnetograph and a H$\alpha$ Lyot filtergraph with Domeless Solar Telescope at Hida Observatory on October 22, 2006.  In H$\alpha$ wing images, we could see many Ellerman bombs (EBs) in the EFR.  Two observation modes, slit scan and slit fixed, were performed with the vector magnetograph, along with H$\alpha$ filtergraph.  Using H$\alpha$ wing images, we detected 12 EBs during slit scan observation period and 9 EBs during slit fixed observation period.  With slit scan observation, we found that all the EBs were distributed in the area where the spatial gradient of vertical field intensity was large, which indicates the possibility of rapid topological change in magnetic field in the area of EBs.  With slit fixed observation, we found that EBs were distributed in the areas of undulatory magnetic fields, both in vertical and horizontal components.  This paper is the first to report the undulatory pattern in horizontal components of magnetic field, which is also evidence for emerging magnetic flux by Parker instability.  These results allow us to confirm the association between EBs and emerging flux tubes.  Three triggering mechanism of EBs is discussed with respect to emerging flux tubes: 9 out of 21 EBs occurred at the footpoints of emerging flux tubes, 8 out of 21 EBs occurred at the top of emerging flux tubes, and 4 out of 21 EBs occurred at unipolar region.  Each case can be explained by magnetic reconnection in the low chromosphere.

\end{abstract}




\section{INTRODUCTION}

  Ellerman bombs (EBs) are short-lived, small-scale bright points observed best in the wings of chromospheric lines, such as H$\alpha$ and Ca {\footnotesize II} H  line.  EBs were first discovered by Ellerman in 1917 \citep{key-1}.  The spectral profile of EBs consists of an absorption core at the H$\alpha$ line center and asymmetric wide emissions in both red and blue wings of H$\alpha$ \citep{key-2,key-3,key-4,key-5,key-6}.

   The mean lifetime of EBs is estimated to be 10-20 minutes \citep{key-9,key-10,key-11}.  The typical size of EBs is of the order of 1{\arcsec} and their shapes are mostly elliptical \citep{key-9,key-10,key-12,key-26}.  The energy of EBs is estimated in the range $10^{25}$-$10^{28}$ ergs \citep{key-5,key-13,key-14,key-12,key-7}.  
   
   It is reported that EBs are associated with chromospheric upflows with a velocity about 6-8 km s$^{-1}$ \citep{key-9,key-6}.  On the other hand, Georgoulis et al. (2002) found that almost all EBs were also accompanied by photospheric downflows with a velocity of 0.1-0.4 km s$^{-1}$.  EBs are also related to H$\alpha$ surges \citep{key-15,key-40}, which are thought to be evidence of magnetic reconnection in the low chromosphere.  Some of chromospheric anemone jets observed with Ca {\footnotesize II} H filters are suggested to be associated to EBs (Shibata et al. 2007).
   
   It is important to study the relationship between EBs and magnetic fields, because EBs generally occur in areas of flux emergence and strong magnetic fields.  In previous studies, it is reported that several EBs appear at the boundaries of well-defined magnetic features, i.e., near magnetic neutral lines \citep{key-16,key-17}, near bipolar emerging fluxes \citep{key-18}, or moving magnetic features \citep{key-10}.  In an emerging flux region (EFR),  EB production is enhanced at the vicinity of sunspots, at areas near magnetic neutral lines and at the boundaries of the neighboring supergranular cells \citep{key-11,key-12}.  Georgoulis et al. (2002) suggested that EBs occur preferentially on separatrix or quasi-separatrix layers in the low chromosphere.  

   Based on observational facts, several triggering models of EBs have been proposed. Most generally accepted one is magnetic reconnection in the low chromosphere \citep{key-14,key-12,key-21,key-19,key-7}.  Chen et al. (2001) confirmed that the magnetic reconnection in the lower solar atmosphere can explain the main characteristics of EBs using a two-dimensional MHD simulation.  On the other hand, Qiu et al. (2000) found that the majority of the H$\alpha$-UV(ultra violet continuum at 1600{\AA}) well-correlated EBs were located at the boundaries of unipolar area, and argued that EBs located at unipolar magnetic areas could be triggered by another mechanism different from magnetic reconnection at low chromosphere.  From topological analysis, Pariat et al. (2004) found that EBs and bald patches were linked by a hierarchy of elongated flux tubes showing aperiodic spatial undulations, whose wavelengths were typically above the threshold of the Parker instability.  These findings led to a resistive emergence model of magnetic flux tubes: EBs are the signature of resistive emergence of undulatory flux tubes provoked by Parker instability.  Isobe, Tripathi, \& Archontis (2007) confirmed the validity of this model by a two-dimensional MHD simulation.    
  
  Our motivation is to test magnetic reconnection model of EBs by analyzing time series data of magnetic field and velocity field in an EFR.  In this paper, we use the high resolution data from spectro-polarimetric observation of NOAA 10917 taken with a vector magnetograph at Domeless Solar Telescope (DST).  Observational information and data reduction are presented in \S 2.  The analysis of morphological characteristics from slit scan observation are described in \S 3.  Temporal evolutions of magnetic field and velocity field around EBs observed in slit fixed observation are shown in \S 4.  Finally in \S 5, we give a discussion and our conclusion about the triggering mechanism of EBs.

\section{OBSERVATION AND DATA REDUCTION}

\begin{figure}  
  \epsscale{.90}
      \plotone{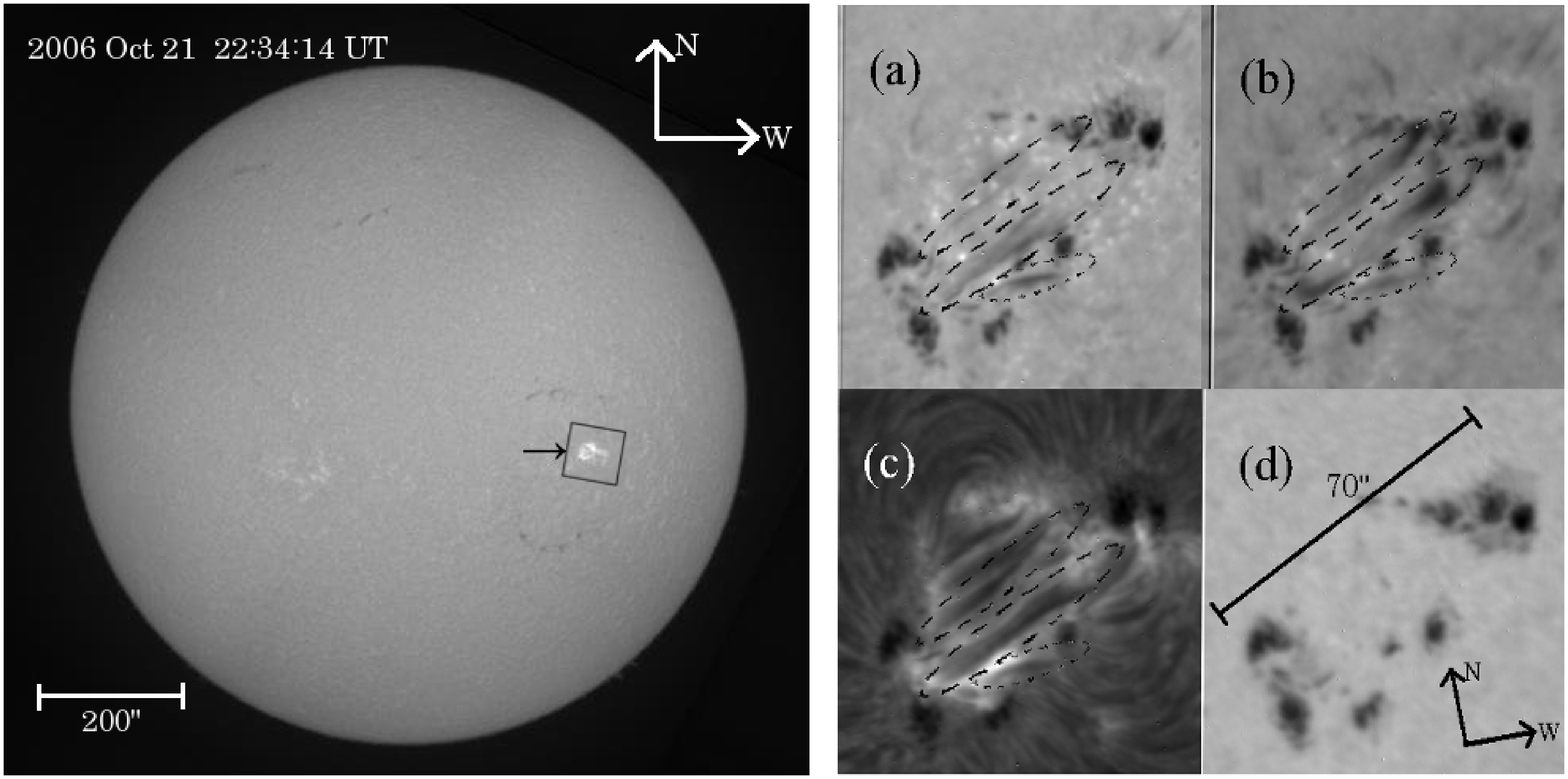}  
 \caption{Left: Full sun image in H$\alpha$ line center on October 21, 2006 taken with SMART.  NOAA 10917 is pointed with an arrow.  The square box indicates the field of view of right panels.  Right: H$\alpha$ images of NOAA 10917 taken with DST.  Four images are taken nearly simultaneously at October 22 00:59 UT.  (a)H$\alpha$ $-$0.8{\AA} (b)H$\alpha$ $+$0.8{\AA} (c)H$\alpha$ line center (d)H$\alpha$ $-$5.0{\AA}.  Three emerging flux tubes (EFTs) which show downward motions at their footpoints and upward ones at their tops are encircled with broken lines.}
  \label{fig:Halpha}
\end{figure}

We observed a $\beta$ sunspot in an active region (AR) NOAA 10917 from October 21, 2006 through October 22 for about 2.5 hours.  Left panel in figure \ref{fig:Halpha} shows a full sun image of October 21 taken with Solar Magnetic Activity Research Telescope (SMART) at Hida Observatory.  With DST, we performed imaging observation of the region in H$\alpha$ on band and off band, and Fe {\footnotesize I} spectroscopic observation with the vector magnetograph.  The AR consisted of two main preceding and following spots, and several small spots (see right panel in figure \ref{fig:Halpha}).  In H$\alpha$ line center images (figure \ref{fig:Halpha}(c)), there were several dark filaments connecting the leading and the following one.  The dark filaments showed red shift at their footpoints and blue shift at their tops, therefore they were rising flux tubes.  Beneath the dark filaments, there appeared numerous EBs in H$\alpha$ wing images (figure \ref{fig:Halpha}(a), (b)).  The heliocentric coordinate of NOAA 10917 at 00:00 UT on October 22 was 5{\degr}S and 33{\degr}W.  The AR was rather stable and in its developing phase during our observation period.  No flare activity during our observation was reported by Space Environment Center.

\subsection{H$\alpha$ filtergram}

10 wavelengths set of H$\alpha$ images, line center, $\pm 0.3${\AA}, $\pm 0.5${\AA}, $\pm 0.8${\AA}, $\pm$ 1.2{\AA}, $-$5{\AA}, were obtained repeatedly with DST from October 21 22:50 UT through October 22 01:27 UT.  In our analysis, we used H$\alpha$ $\pm$0.8{\AA} images for identification of EBs.  The pixel size of H$\alpha$ filtergram was 0{\arcsec}.256, and the field of view (FOV) was 261{\arcsec} x 261{\arcsec}.  The time cadence of each wavelength frame was $\sim$40 seconds.  All the data were corrected by dark current subtraction and flat fielding.

\subsection{Vector magnetograph}

Along with H$\alpha$ filtergrams, Fe {\footnotesize I} spectra were observed using a vector magnetograph (VMG).  The spectral FOV includes both absorption lines of Fe {\footnotesize I} 6301.5{\AA} and Fe {\footnotesize I} 6302.5{\AA}.  The spectral resolution is 0.0083{\AA} pixel$^{-1}$ at 6302.5{\AA}.  The slit length was 140{\arcsec}.8 and the width was 0{\arcsec}.275.  VMG consists of a rotating waveplate, three Wollaston prisms, and a high-dispersion spectrograph (Kiyohara et al. 2004).  During one rotation of the waveplate, a pair of spectral images in orthogonally polarized states were taken at every 22.5{\degr} interval.  It took $\sim$30 seconds for one rotation of the waveplate.  16 pairs data were demodulated into the Stokes vector I, Q, U, V with the method described in Kiyohara et al. (2004).  The calculated Stokes I, Q, U, V were inverted to parameters such as magnetic field intensity, magnetic inclination, magnetic azimuth, and line-of-sight velocity by ASP (Advanced Stokes Polarimeter) inversion code (see Skumanich \& Lites (1987)).  The ASP inversion technique adopts a hypothesis of a Milne-Eddington atmosphere, so that parameters are assumed to be constant along line-of-sight direction.  Treatment of inversion was applied only in  the area whose polarization degree was larger than 0.5\%.  Noise levels of the observed quantities, which were estimated from the temporal fluctuations at a relatively stable point near the center of the AR, were $\pm$50 Gauss for magnetic field intensity, and $\pm$3{\degr} for inclination and azimuth angle.

We performed two kinds of observation mode using VMG.  One was "slit scan" and the other was "slit fixed."  In the slit scan mode, we scanned the AR by 2{\arcsec} intervals for 46 positions, so that we could get one map with 92{\arcsec} x 140{\arcsec}.8 FOV.  In the slit fixed mode, we tracked the AR manually aligning the slit fixed along the axis of the AR, where EB production was active.  The exposure time and the time cadence were 1500 ms and $\sim$30 seconds for both observation modes.  First, slit fixed observation was done from October 21 23:03 UT to October 22 00:53 UT.  Next, slit scan observation was done from October 22 01:00 UT to 01:25 UT. 

All spectral data were corrected by dark current subtraction and flat fielding.  Slit fixed data were co-aligned with reference to the position of the main spots on continuum images.  We performed box car smoothing of 5 frames ($\sim$2.5 minutes) on the slit fixed data to reduce high frequency noise.  For time series data of Doppler velocity, we applied lowpass filter ($\leq$3 mHz) in order to eliminate the effect of photospheric five minutes oscillation.  To get the physical insights into the magnetic configuration more easily, we corrected for geometric projection effect to the vector magnetic field.  To be more precise, the magnetic inclination was converted to the tilt angle from local normal to the solar surface, and the azimuth was converted to the tilt angle measured from the slit direction on local horizontal plane.   

\subsection{Identification of EBs}

As was stated in \S 1, EBs are characterized by their dark core and bright wing enhancements in H$\alpha$.  To identify EBs, we calculated the contrast $I_{c}(\mbox{\boldmath $x$},t)=[I(\mbox{\boldmath $x$},t)-I_{0}(t)]/I_{0}(t)$ on H$\alpha$ $\pm$0.8{\AA} images.  Here $I(\mbox{\boldmath $x$},t)$ is the intensity of a pixel at position $\mbox{\boldmath $x$}$ taken at time \textit{t}, and $I_{0}$ is the mean intensity of the entire frame at time \textit{t}.  We adopted an empirical threshold value of $I_{c}=$1.15 for EBs, that is, EBs are the regions where intensity is more than 15\% brighter than the average.  This value is consistent with earlier works, 5\%$-$30\% in Georgoulis et al. (2002) and 14\% in Pariat et al. (2007).  We made sure that they did not show any brightness enhancements at the positions of EBs in H$\alpha$ center images.  

\begin{figure}
  \epsscale{.90}
    \plotone{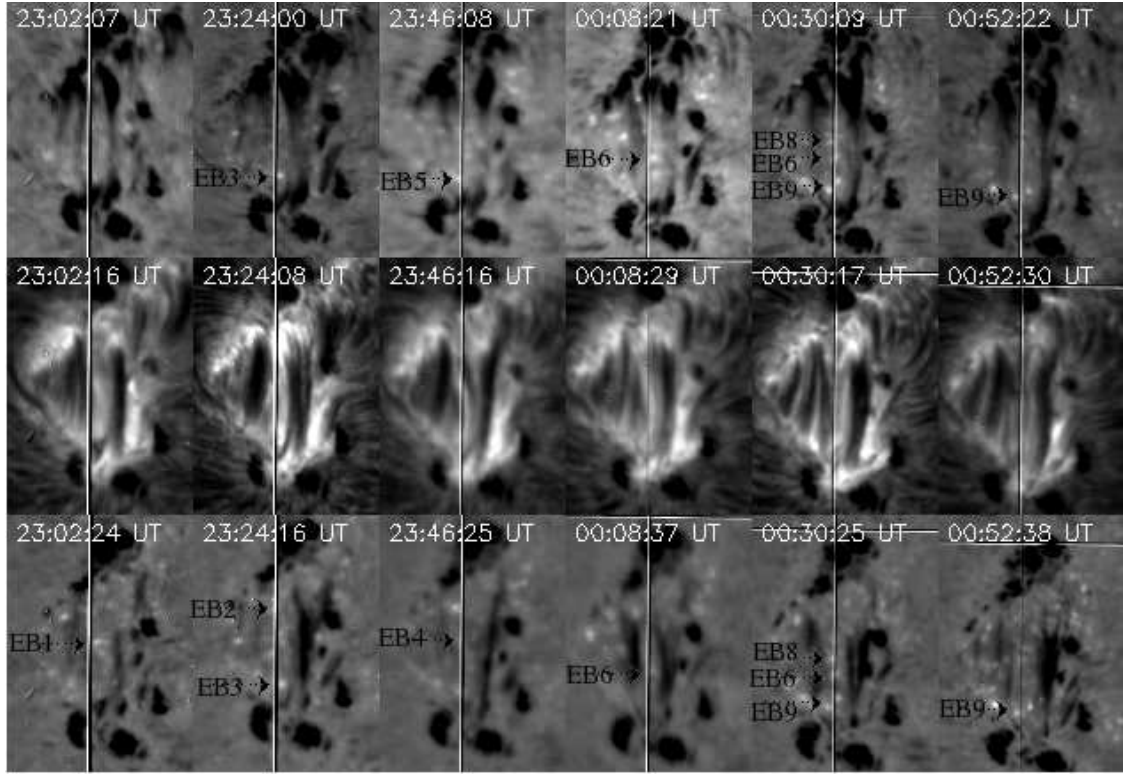}  
  \caption{H$\alpha$ wing and center images taken during the slit fixed observation.  The upper 6 panels are H$\alpha$ +0.8{\AA} images, in which downward moving regions appear dark.  The middle 6 panels are H$\alpha$ center images.  Lower 6 panels are H$\alpha$ $-$0.8{\AA} images, in which upward moving regions appear dark.  The position of each identified EBs are pointed with arrows.}\label{fig:Ha_EFT_fixed}
\end{figure}

In order to know the EFR magnetic conditions around EBs, we analyzed only such EBs that were adjacent to the VMG slit.  As a result, 12 EBs were detected during slit scan observation, and 9 EBs during slit fixed observation.  Figure \ref{fig:Ha_EFT_fixed} shows temporal series of H$\alpha$ wing and line center images and the position of 9 EBs during slit fixed observation period.  In figure \ref{fig:Ha_EFT_fixed}, the appearances of the dark filaments and their velocity distributions change rapidly in time.  This may be due to new emerging flux tubes (EFTs) which pressed the dark filament from below the photosphere.  Among 9 EBs during slit fixed observation, 7 EBs were observed from their birth to their death, another EB already existed when the observation started, and 1 EB did not yet decay when the observation stopped.

\section{RESULTS OF SLIT SCAN OBSERVATION}

\begin{figure} 
  \epsscale{.60}
   \plotone{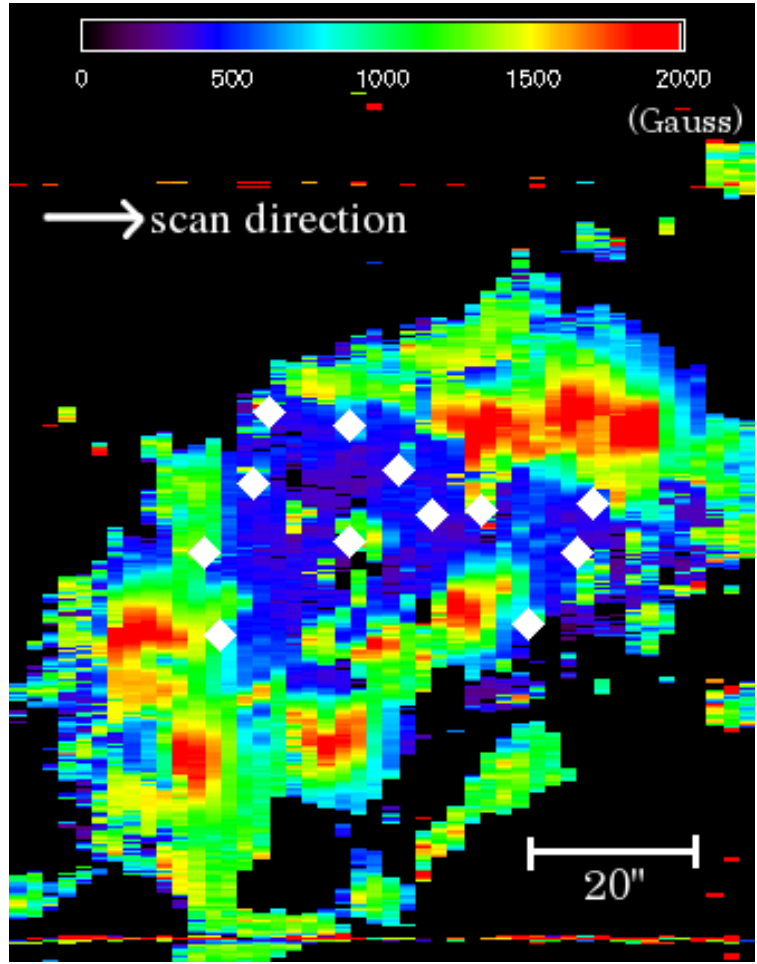}  
  \caption{Spatial distribution of magnetic field intensity (slit scan).  White diamonds show the positions of observed EBs.}\label{fig:scan_intensity}
\end{figure}

Figure \ref{fig:scan_intensity}-\ref{fig:scan_velocity} show results obtained from slit scan observation.  The slit scanning direction was in transversal, from left to right in each figure.  The positions of 12 identified EBs are indicated by yellow or white diamonds.  In this section, we mainly describe the morphological characteristics of the AR and preferential locations of EBs.  

\subsection{Magnetic field}

\begin{figure} 
 \epsscale{.60}
   \plotone{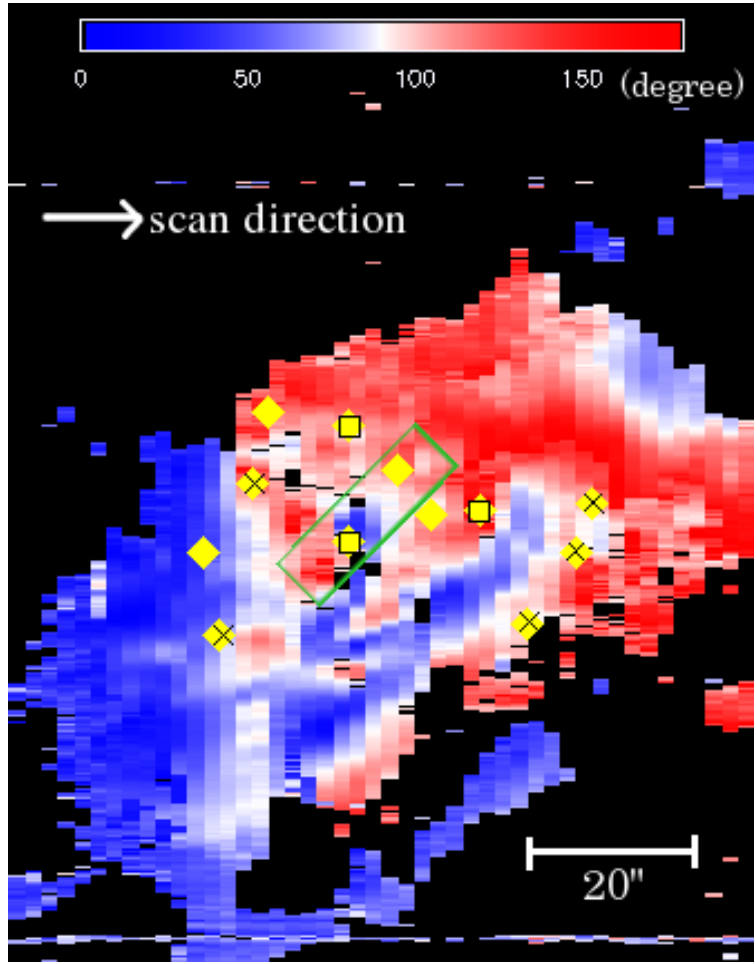}  
 \caption{Field inclination with respect to local normal (slit scan).  Yellow diamonds show the positions of observed EBs.  Cross signs indicate EBs at the "top", and square signs indicate EBs at the "dip".  The magnetic configuration in green rectangle area is given in figure \ref{fig:green}.}\label{fig:scan_inclination}
\end{figure}

\begin{figure}  
  \epsscale{.60}
   \plotone{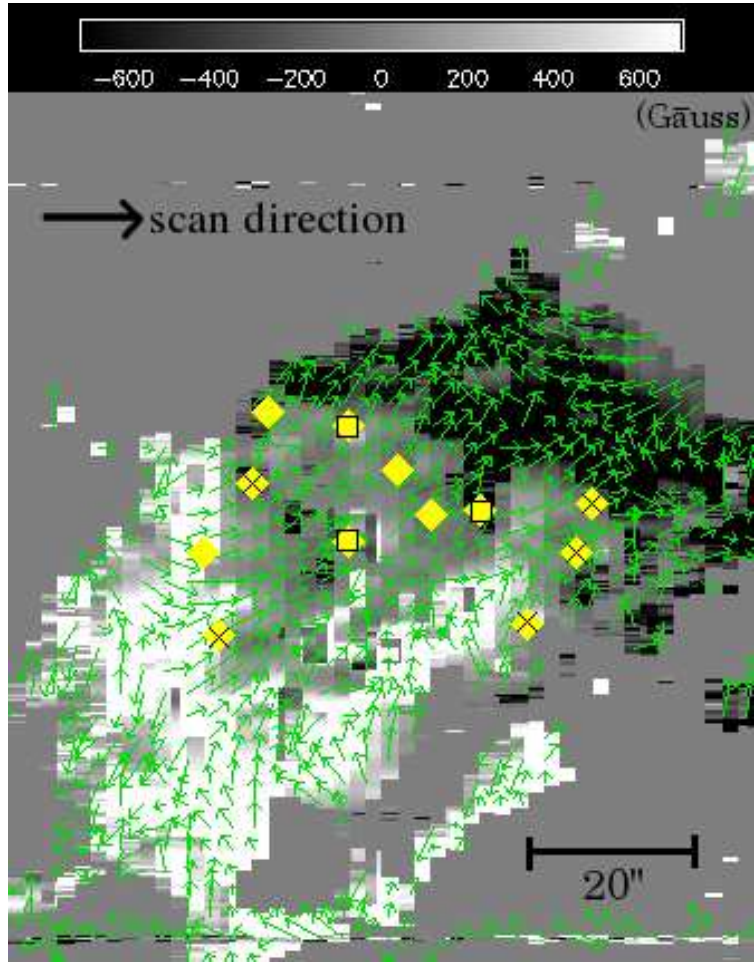}  
  \caption{Vertical component of magnetic field intensity (background gray scale) and horizontal magnetic field vectors (slit scan).  The length of each arrow is proportional to log-scaled horizontal components of magnetic field.  Yellow diamonds show the positions of observed EBs.  Cross signs indicate EBs at the "top", and square signs indicate EBs at the "dip".}\label{fig:scan_magline}
\end{figure}

Figure \ref{fig:scan_intensity} shows the distribution of magnetic field intensity.  Small spots had large field intensity $\sim$2000 Gauss (see figure \ref{fig:Halpha} for reference).  EBs were distributed at regions with field intensity 500-1000 Gauss.  Field inclination angle shown in figure \ref{fig:scan_inclination} means the tilt angle from local normal to the solar surface, and the background gray scale in figure \ref{fig:scan_magline} shows vertical components of magnetic field, i.e., (field intensity)$\times$cos(field inclination).  After the correction for the projection effect, we solved 180{\degr} ambiguity of azimuthal angle with a hypothesis that the magnetic lines were similar to those of a dipole between the two main spots, because NOAA 10917 was a simple dipole in overview.  Green arrows in figure \ref{fig:scan_magline} indicate horizontal vectors of magnetic lines whose lengths are proportional to log-scaled horizontal components of magnetic field at each point.  As is seen in the background image of figure \ref{fig:scan_magline}, that EB production areas showed large spatial gradient of vertical field intensity.  This suggests the possibility of rapid topological change of magnetic field in the area of EBs, probably due to new EFTs.

We found that most EBs were distributed at regions whose inclination angles $\sim$90{\degr}, that is, near magnetic neutral line.  We found 8 out of 12 EBs near magnetic neutral lines (square and cross signs in figure \ref{fig:scan_inclination}-\ref{fig:scan_velocity}), while 4 EBs in unipolar region.  Here, we introduce two classifications of neutral lines, "top" and "dip."  They are defined by the sign of the field line curvature \citep{key-19}.  "Top" is a region of the photosphere ($z=0$) where the field line curvature is negative,
\begin{equation}
B_{z}=0 \ \ \textrm{and}\ \  \mbox{\boldmath{$B \cdot \nabla$}}  B_{z} < 0
\end{equation}
while "dip" is a region where the field line curvature is positive, 
\begin{equation}
B_{z}=0 \ \ \textrm{and}\ \   \mbox{\boldmath{$B \cdot \nabla$}} B_{z} > 0
\end{equation}
"Top" is sometimes referred to $\Omega$ loop, and "dip" is U loop.  Then the 8 EBs found on neutral lines are classified into 5 "top" EBs and 3 "dip" EBs.  In figure \ref{fig:scan_inclination}-\ref{fig:scan_velocity}, the cross symbols mean "top" and the squares mean "dip."

\subsection{Doppler velocity}

\begin{figure}
  \epsscale{.60}
    \plotone{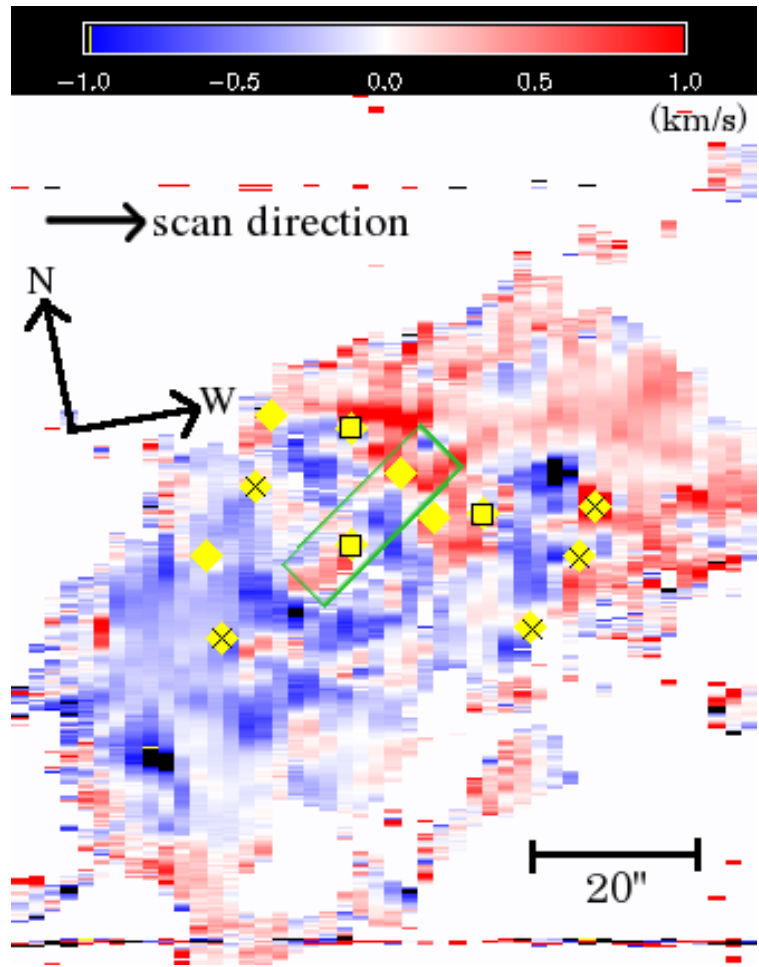}  
  \caption{Line-of-sight velocity (slit scan).  Negative velocity means vertically upward or horizontally eastward motion, and positive velocity means vertically downward or horizontally westward motion.  Yellow diamonds show the positions of observed EBs.  Cross signs indicate EBs at the "top," and square signs indicate EBs at the "dip".  The magnetic configuration in green rectangle area is given in figure \ref{fig:green}.}\label{fig:scan_velocity}
\end{figure}

Figure \ref{fig:scan_velocity} shows Doppler velocity field, or line-of-sight velocity field.  Red regions have receding velocity from us and blue regions have approaching velocity to us.  The effect of solar rotation velocity is already subtracted.  There was no  significant velocity field with $|v|>2$ km s$^{-1}$ around this AR.  We assumed that the observed Doppler velocity consisted of large scale flow field and localized flow field.  In large scale view, blue trend in the east side and red trend in the west side were considered to be horizontal flux segregation motion of the EFR or supergranular horizontal diverging flow (Bernasconi et al. 2002; Kozu et al. 2006; Magara 2006).  Because the target's position was in the west side, the horizontal segregation motion is observed as blue in the east side and red in the west side.  Using SOHO/MDI full disk magnetogram images, we estimated the horizontal velocity around this AR based on the local correlation tracking method.  The averaged horizontal velocity over our observational period was 0.16 km s$^{-1}$, diverging from the center to east and west direction.  Then the effective component of horizontal velocity against the line-of-sight velocity was less than 0.1 km s$^{-1}$.

\begin{figure} 
  \epsscale{.60}
    \plotone{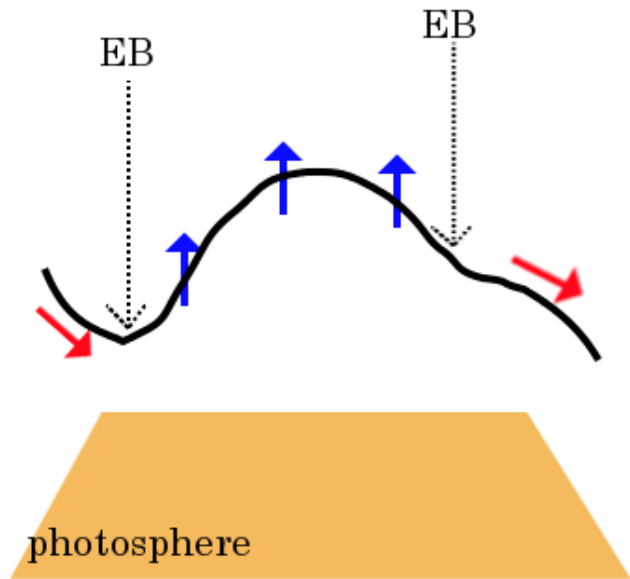}  
  \caption{Plausible configuration of magnetic field lines in the area enclosed with a green rectangle in figure \ref{fig:scan_inclination} and figure \ref{fig:scan_velocity}.  Blue and red arrows indicate the direction of plasma mass flow.}\label{fig:green}
\end{figure}

Let us look at the localized relation between EB positions and Doppler map in figure \ref{fig:scan_velocity}.  The EBs were mostly located on the boundary of red/blue Doppler velocity.  This result also proved the existence of EFTs.  One example is shown in the areas enclosed with a green rectangle in figure \ref{fig:scan_inclination} and \ref{fig:scan_velocity}.  There existed an $\Omega$ shaped magnetic field line.  Its top showed blue shift, or upward motion of the magnetic line and its footpoints showed red shift because of the plasma sliding down along the loops.  Thus the observed local Doppler variation around the EBs can be interpreted as was sketched in figure \ref{fig:green}.  So the estimated amplitude of horizontal velocity ($<$ 0.1 km s$^{-1}$ from SOHO/MDI magnetogram images) and the results of earlier works (Lites et al. 1998; Kozu et al. 2006) allow us to suppose the localized line-of-sight velocities were predominantly contributed by the vertical motion.
  

\section{RESULTS OF SLIT FIXED OBSERVATION}

\begin{figure}  
  \epsscale{.60}
    \plotone{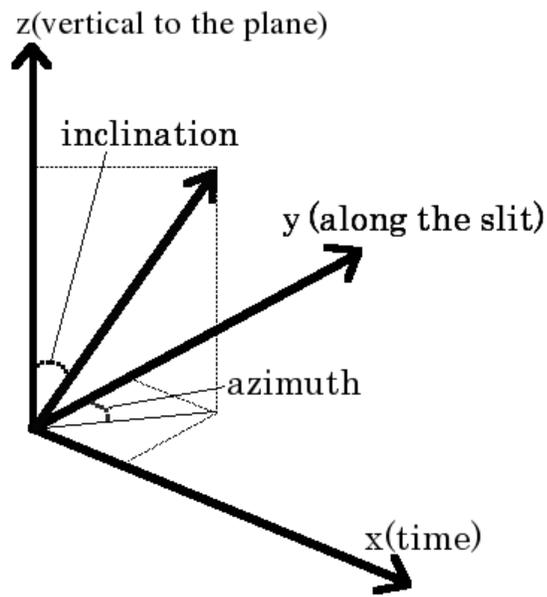}  
  \caption{Coordinate system used in slit fixed observation results.}\label{fig:coordinate}
\end{figure}

\begin{figure}  
  \epsscale{.80}
    \plotone{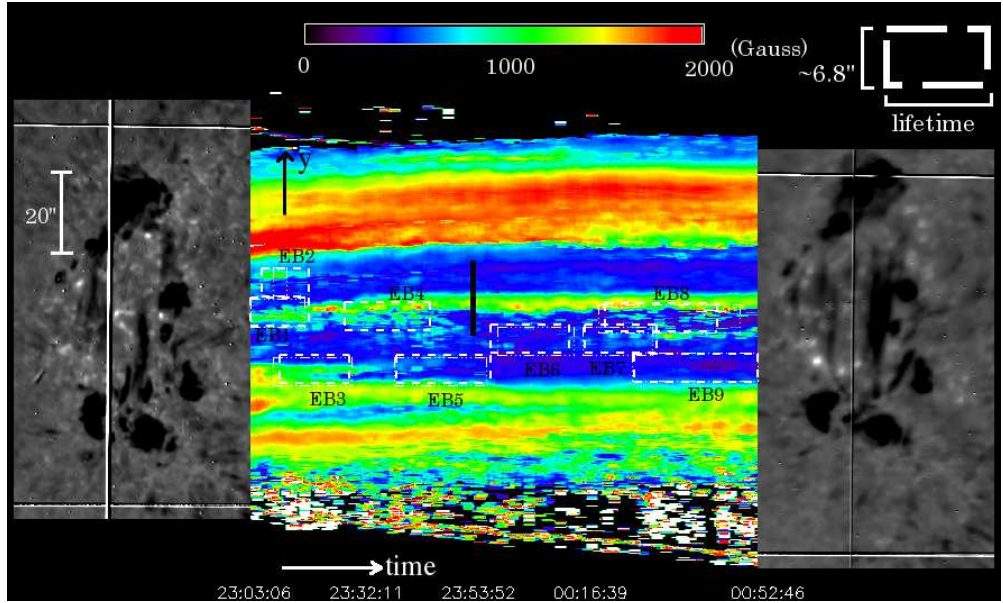}
  \caption{Magnetic field intensity variation with time (slit fixed).  Times are shown in UT.  The rectangles are indicators of EB positions.  The horizontal length of the rectangle indicates its lifetime, and the vertical width covers a zone of the EB position  $\pm$3{\arcsec}.4.  Two H$\alpha$ $-$0.8 {\AA} images are given for references at the beginning and the end of the observation.  Plot along the black solid line is shown in figure \ref{fig:emerging}(a).}\label{fig:intensity}
\end{figure}

\begin{figure}  
  \epsscale{.80}
    \plotone{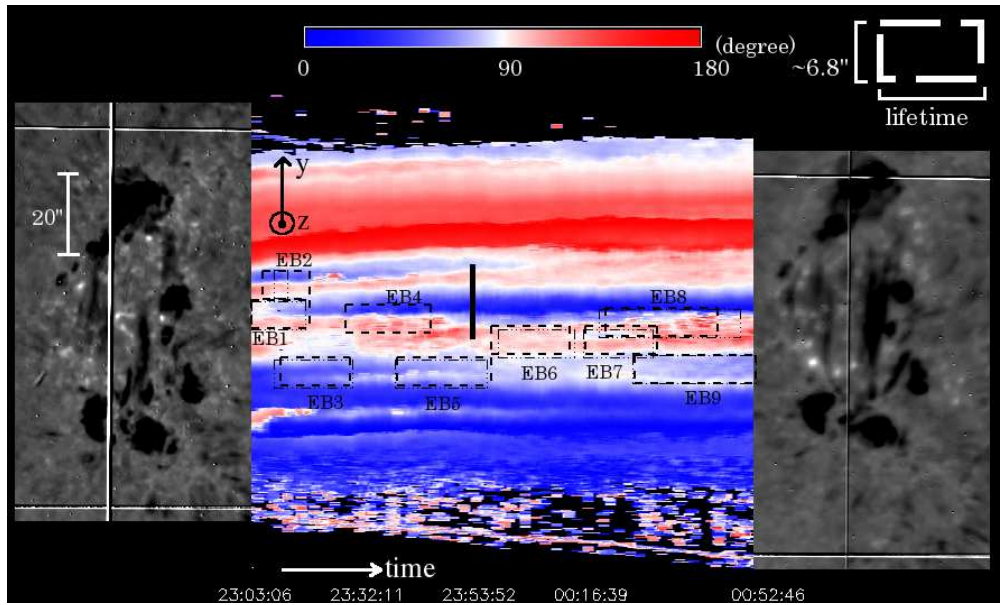}   
  \caption{Magnetic inclination variation with time (slit fixed).  The others are the same as in Fig. \ref{fig:intensity}.  Plot along the black solid line is shown in figure \ref{fig:emerging}(b).}\label{fig:inclination}
\end{figure}

\begin{figure} 
  \epsscale{.80}
    \plotone{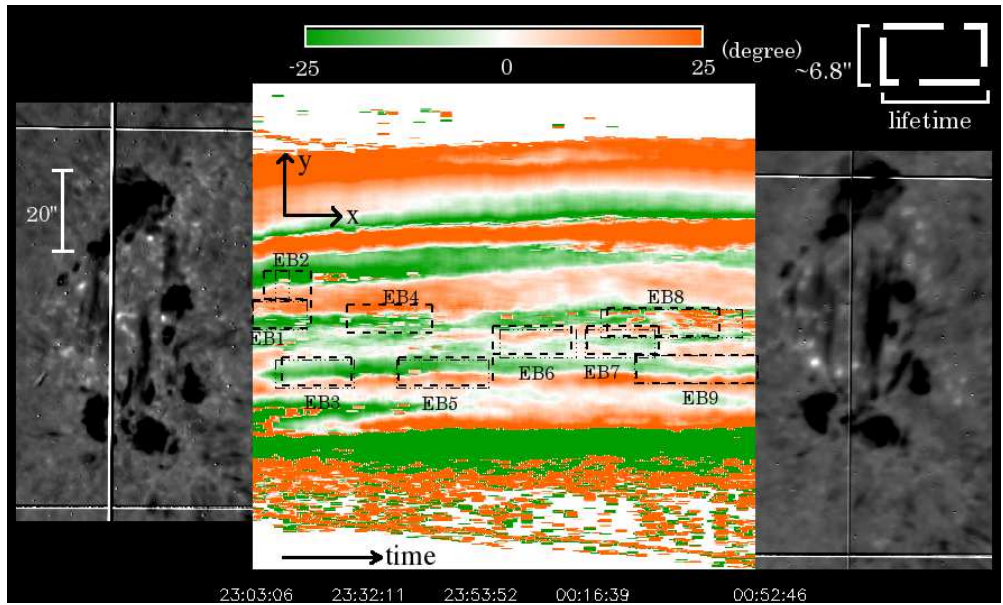}     
  \caption{Magnetic azimuth variation with time (slit fixed).  The others are the same as in Fig. \ref{fig:intensity}.}\label{fig:azimuth}
\end{figure}

\begin{figure} 
  \epsscale{.80}
    \plotone{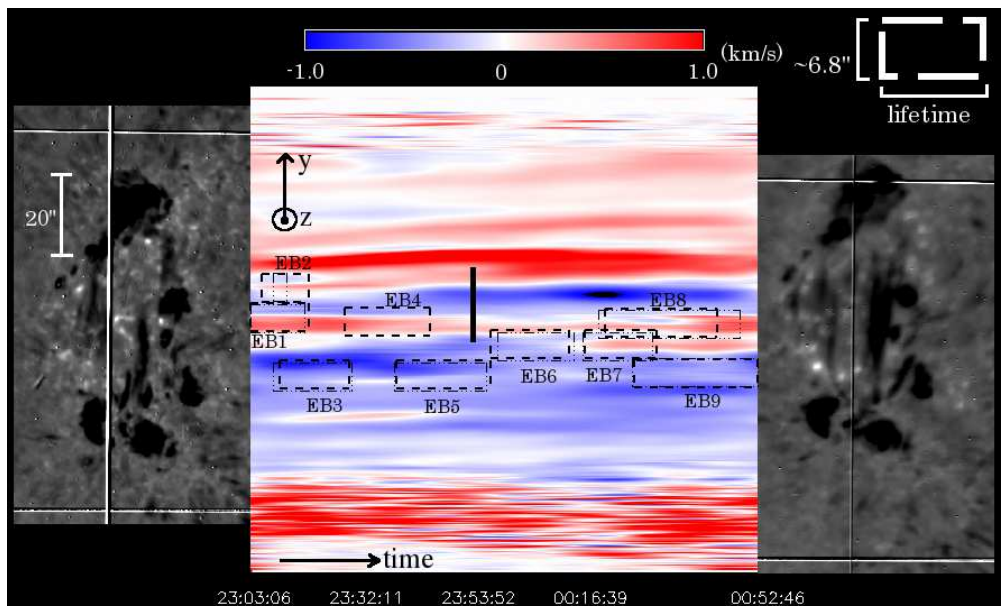}  
  \caption{Doppler velocity variation with time (slit fixed).  Negative means blue shift, and positive means red shift.  The others are the same as in Fig. \ref{fig:intensity}.  Plot along the black solid line is shown in figure \ref{fig:emerging}(c).}\label{fig:velocity}
\end{figure}

Next, we proceed to the results obtained from slit fixed observation.  Figure \ref{fig:coordinate} explains the coordinate system used in figure \ref{fig:intensity}-\ref{fig:velocity}.  The left and right H$\alpha$ images are respectively H$\alpha$ $-$0.8{\AA} images taken at the beginning of the observation and at the end.  For example, in the left H$\alpha$ panel, we can see a bright patch near the center of the slit, which is marked as EB1 in the middle panel.  The middle panels show the temporal variation of physical parameters on the slit.  The dashed and dash-dotted rectangles are indicators of EB positions.  The dashed rectangles show EBs found by H$\alpha$ $-$0.8{\AA} images, and the dash-dotted ones show EBs found by H$\alpha$ +0.8{\AA} images.  As is demonstrated at upper right in each figure, the horizontal length of the rectangle indicates its lifetime, and the vertical width covers a zone of the EB position $\pm$3{\arcsec}.4.  The mean lifetime of the 9 detected EBs was $\sim$20 minutes, which is comparable to those of previous works.  As is in the slit scan observation, 180{\degr} ambiguity of azimuthal angle is resolved with a hypothesis that the magnetic lines were similar to those of a simple dipole.

The slit fixed figures \ref{fig:intensity}-\ref{fig:velocity} appear to illustrate another important EB characteristic, i.e., that EBs appear to occur and recur at preferential locations.  Because the slit's position was fixed along the axis of the AR, we can easily classify into three groups of EBs that faded and reappeared at nearly the same locations.  These groups are (EB3, EB5, EB9), (EB6, EB7), and (EB1, EB4, EB8).  This is a nice result which shows occurrence and recurrence of EBs with temporarily varying underlying magnetic conditions.

\subsection{Magnetic field intensity}

Figure \ref{fig:intensity} shows temporal evolution of magnetic field intensity on the slit.  As we mentioned in \S 3.1, EBs were distributed at regions with 500-1000 Gauss magnetic field intensity.  Magnetic field intensity near EBs was evidently increasing, or at least changing in time.  If a new EFT makes an appearance from beneath the photosphere, it presses the pre-existing magnetic field and magnetic intensity gradually increases.  Therefore strong magnetic intensity area can be interpreted as an EFT production site.  Except for EB6, EB7, and EB9, the other 6 EBs showed signatures of increasing field intensity.

\subsection{Magnetic inclination}

\begin{figure} 
  \epsscale{.50}
    \plotone{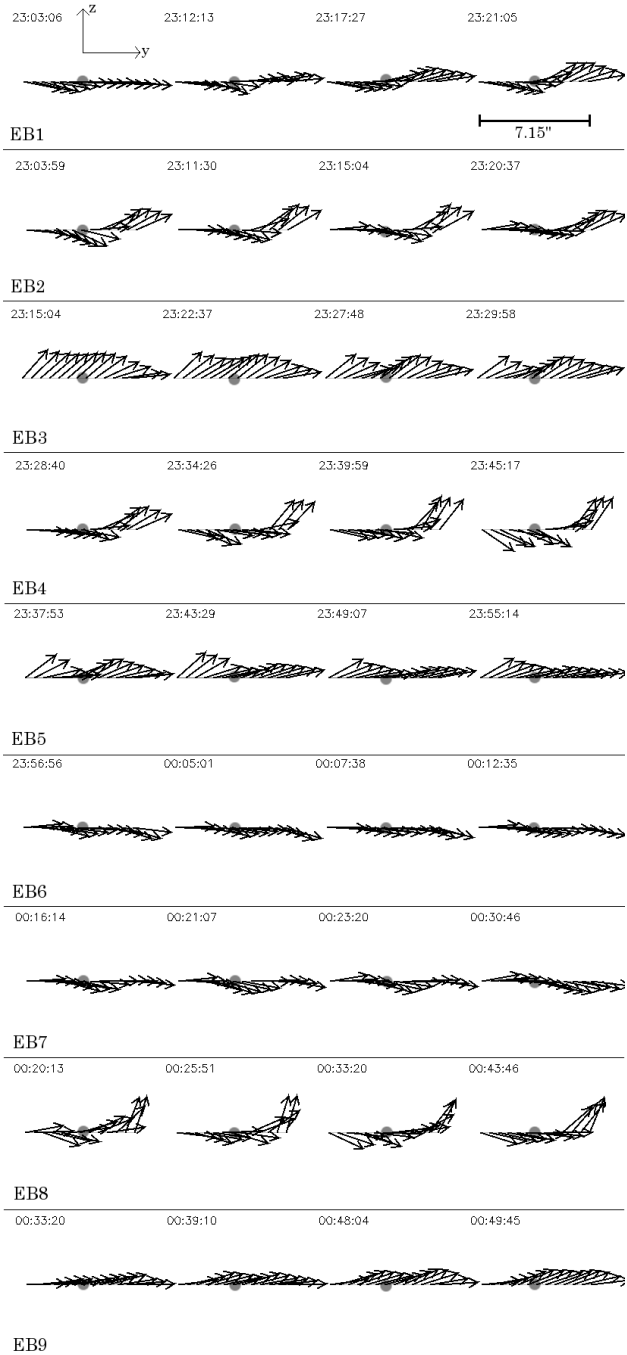}  
  \caption{Temporal evolution of magnetic field inclination around each EB.  Time for each frame is shown at upper left in UT.  Circle symbol at the center of each figure indicates the position of the EB.}\label{fig:inclination_line}
\end{figure}

In figure \ref{fig:inclination}, temporal evolution of magnetic field inclination with respect to local normal (+z) is shown.  We made a more detailed picture of temporal variation of inclination angle around each EB in figure \ref{fig:inclination_line}.  In figure \ref{fig:inclination_line}, arrows indicate the magnetic vectors (y,z)=(cos(inclination), sin(inclination)) with unit length.  As is the same in \S 3.1, almost all the EBs were distributed near magnetic neutral lines.  

According to the definition explained in \S 3.1, EB1, EB2, EB4, EB6, EB7 and EB8 belonged to "dip."  No "top" EB was found.  And EB3, EB5, and EB9 belonged to neither of them but on unipolar region.  Although EB3 and EB5 did not lie in magnetic neutral line but stay in unipolar region, their distributions of inclination angles showed apparent dip configuration, which is clearly seen in figure \ref{fig:inclination_line}.  We shall call this kind of configuration as "local dip" from now on.

At the locations of EB1 and EB3, we found temporal formation of magnetic dip structure.  At first, the distribution of magnetic field was uniform along the slit.  As the time passed, inclination of field lines evolved such that the magnetic topology took the dip structure in the photosphere.  It is interesting that H$\alpha$ wing brightenings already started while the magnetic field lines were still in uniform state in the photosphere, which may indicate the magnetic dips were already formed in the upper layers than the Fe {\footnotesize I} 6302.5{\AA} line formation heights and started to produce an EB by magnetic reconnection mechanism.

In the central part of the active region, there was a clear undulatory pattern, which is seen as periodically blue and red regions along the slit in figure \ref{fig:inclination}.  All the EBs were distributed in this undulatory field line area.  This is supporting evidence for the resistive emergence model of Pariat et al. (2004).

\subsection{Magnetic azimuth}

Figure \ref{fig:azimuth} shows temporal evolution of magnetic field azimuth on local horizontal plane.  Azimuthal angle is the tilt angle measured from the slit direction (+y).  Magnetic lines in orange colored region are inclined to right (+x) direction and magnetic lines in green colored region are inclined to left ($-$x) direction.  Azimuthal angles between the two main spots mostly took value ranging from $-$20{\degr} to +20{\degr}.  Like inclination angle, azimuthal angle showed a clear undulatory pattern.  Two plots of inclination and azimuth angle along the slit are shown in figure \ref{fig:undulation}.  The four cross signs are indicators of EB generation sites.  It is clearly seen that the magnetic field around EBs were undulatory, and that EBs were located in the region where the inclination of magnetic lines was $\sim$90{\degr} and their azimuth angles changed their signs.  The characteristic wavelength of the undulation pattern exceeded 5\arcsec, that is, longer than 4000km.  This result tells us an important fact, because 4000 km is the characteristic wavelength of Parker instability at the photospheric layer. As long as we know, this is the first to report an undulatory pattern in azimuthal angles around EBs.

\subsection{Emerging flux tubes}

\begin{figure}  
  \epsscale{.80}
    \plotone{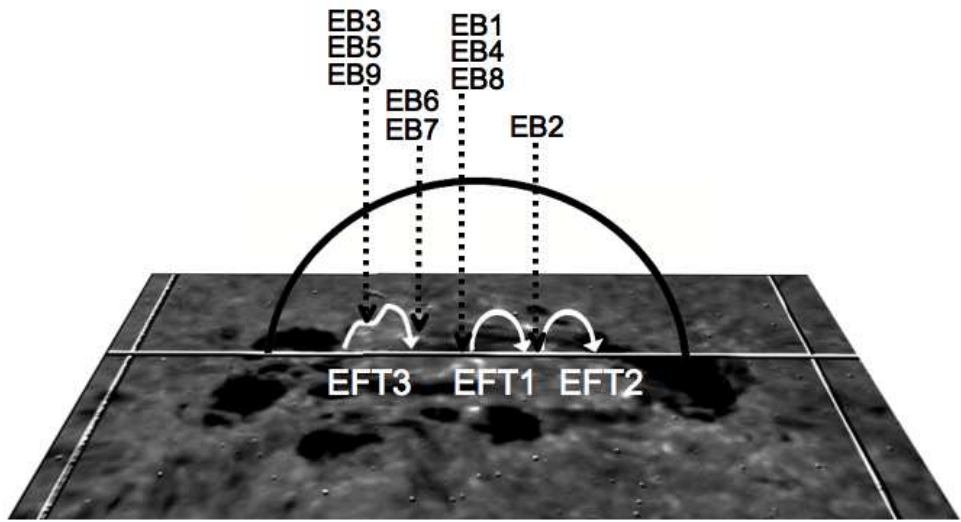}   
  \caption{Position of three EFTs and EBs.  The ground figure is the same as the left H$\alpha$ image shown in figure \ref{fig:intensity}-\ref{fig:velocity}.}\label{fig:EFposition}
\end{figure}

\begin{figure} 
   \epsscale{.50}
    \plotone{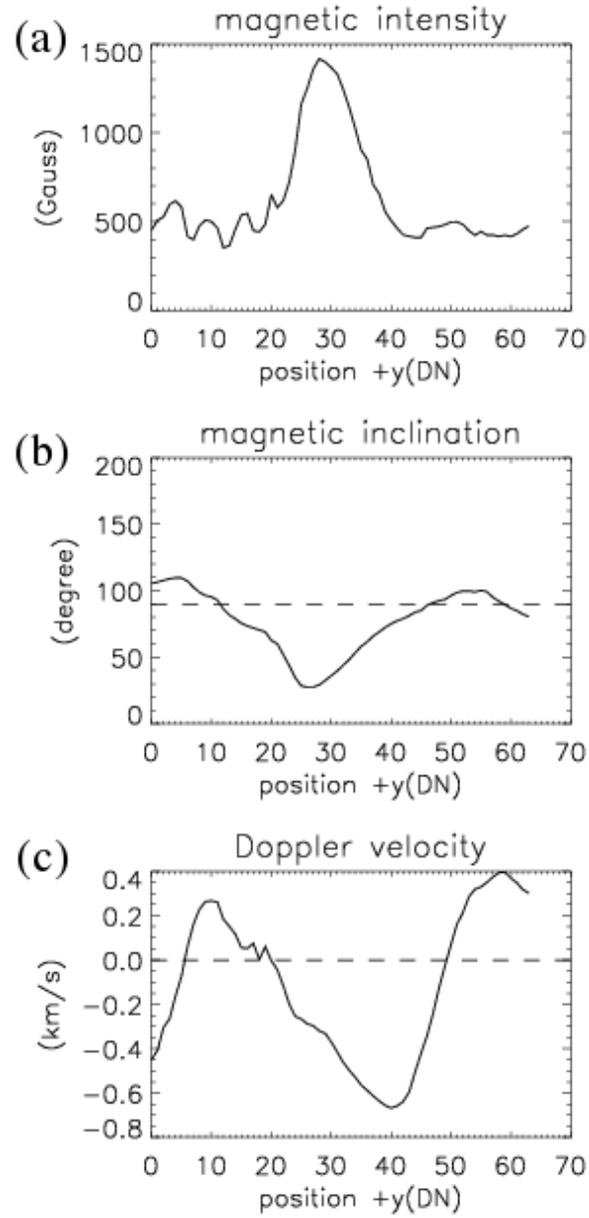}    
  \caption{Three kinds of plots of EFT1 located along a black solid line in figure \ref{fig:intensity}, \ref{fig:inclination} and \ref{fig:velocity}.  (a)magnetic field intensity (b)magnetic inclination (c)Doppler velocity (Negative is upflow and positive is downflow.)}\label{fig:emerging}
\end{figure}

In the observed AR, there were signatures of many EFTs in the photosphere.  
The preferential position of EBs showed an undulatory pattern in their magnetic field components, probably due to new EFTs triggered by Parker instability.  We identified three small EFTs during the slit fixed observation.  Figure \ref{fig:EFposition} shows the positions of three EFTs and EBs.  The identification of an EFT was done by examining the distribution of magnetic field intensity, inclination, and Doppler velocity.  Figure \ref{fig:emerging} shows three kinds of plots, magnetic field intensity, inclination and Doppler velocity along the black solid line in figure \ref{fig:intensity}-\ref{fig:velocity}.  The field line showed $\Omega$ shape in its inclination, the field intensity was increasing, and Doppler velocity changed its sign across the neutral line.  Blue shifted area was found near the top of the $\Omega$ loop field lines.  Blue shifted area can be interpreted as vertically upward motion as was suggested in \S 3.2.  At both sides of the footpoints, there existed red shifted, i.e., downward moving areas.  We think these downflows were due to the motion of plasma dragged from the top of the $\Omega$ loop along field lines.  Then we conclude this area as an EFT, and call this EFT1.  EFT1 was located near EB1 and started emergence from about 23:10 UT and had not yet decayed at the end of our observation.  Second one (EFT2) was located near EB2.  It started emergence before the observation started and decayed at about 00:00 UT.  EFT2 also showed magnetic intensity enhancement and Doppler blue shift $\sim$0.1 km s$^{-1}$ at the top, and Doppler red shift $\sim$1.0 km s$^{-1}$ at its footpoints.  Third one (EFT3) made an appearance near EB3.  It started emergence at about 23:20 and had not yet decayed at the end of the observation.  EFT3 induced magnetic field intensity enhancement at its early phase and Doppler blue shifted velocity $\sim$1.0 km s$^{-1}$.  However unlike EFT1 and EFT2, EFT3 did not have red shifted areas at its footpoints.    
 



\section{DISCUSSION}

\subsection{Schematic models}

With slit scan observation, we found that 8 out of 12 identified EB production areas had large spatial gradient of vertical field components.  This indicates the existence of new EFTs.  Doppler velocity map in figure \ref{fig:scan_velocity} shows that almost all the EBs had downward motion adjacent to upflow, which is also a signature of new EFTs.  Thus we confirmed that EBs have a strong relation to the emergences of EFTs with slit fixed observation.  As for the residual 4 EBs found on unipolar region, we also try to interpret by the emergence of EFTs and magnetic reconnection, as will be discussed below.

During the slit fixed observation period, we identified three EFTs (figure \ref{fig:EFposition}).  Each EFT invoked a gradual enhancement of magnetic field intensity.  In regions where magnetic fluxes emerge continuously, new EFTs compress the pre-existing fluxes and the flux density increases.  If oppositely directed or at least sheared magnetic lines are compressed and approaches each other, a magnetic reconnection occurs.  It is widely accepted that EBs are a signature of magnetic reconnection at the low chromosphere.  Therefore, it is plausible that magnetic field intensity increases at the location of an EFT and an EB.  

\begin{figure} 
  \epsscale{.35}
   \plotone{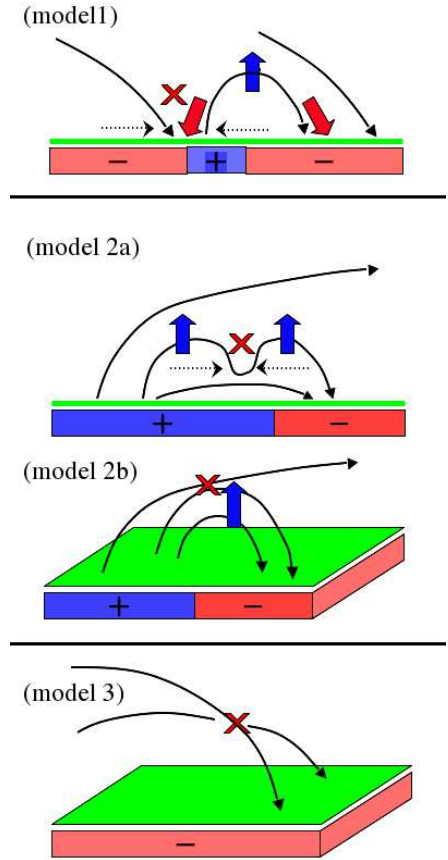} 
  \caption{Three different mechanisms of EB triggering.  (model 1) EB mechanism at the footpoints of EFTs. (model 2) EB mechanism at the top of EFTs. (model 3) EB mechanism at unipolar region.  The layer at the footpoint of vectors indicates the formation height of Fe {\footnotesize I} 6302.5{\AA}.  X points are reconnection points.  Lower rectangles with minus symbol mean that inclination angle is larger than 90{\degr} and rectangles with plus symbol mean that inclination angle is less than 90{\degr}.  Thick arrows show Doppler velocities; blue arrows for upflow and red arrows for downflow.  Broken line arrows written in model 1 and model 2a means horizontal converging flow that may exist.}\label{fig:model}
\end{figure}

In figure \ref{fig:model} we show schematically how our data are interpreted by magnetic reconnection.  Model 1, 2, and 3 show EB mechanisms with different magnetic configurations: Model 1 for footpoint of an EFT, model 2 for top of an EFT, and model 3 for unipolar region.  According to the definition of magnetic lines in \S 3.1, a "dip" EB is explained by model 1, a "top" EB is explained by model 2, and a unipolar by model 2 or model 3.  Due to the coarse scan step and lack of temporal information, we can not clarify that 4 unipolar EBs found on the slit scan observation were located at the top of an EFT or not, and so 4 unipolar EBs can be explained either of model 2 or model 3.  In summary, the identified EBs are classified as follows; (EB1, EB2, EB4, EB6, EB7, EB8) and 3 "dip" EBs of slit scan belonged to model 1.  (EB3, EB5, EB9) and 5 "top" EBs of slit scan belonged to model 2.  Finally, 4 unipolar EBs of slit scan belonged to model 2 or model 3.

In model 1, new EFTs appear from below the pre-existed uniform magnetic field.  There will be produced an anti-parallel layer at the low chromosphere, just above the photospheric "dip" region, and reconnection occurs at X point.  When reconnection occurs, there should exist converging flow across anti-parallel field lines.  This speculative flow is shown in figure \ref{fig:model} with broken line arrows.  We could not detect this speculative horizontal flow in our observation, probably due to their small amplitude of velocity, compactness of spatial extension of the velocity field, and etc.


Model 2 explains the triggering mechanism of an EB on top of an EFT.  Model 2a is the case for local dip configuration, which accounts for EB3 and EB5.  Like in model 1, horizontal converging flow may exist across the local dip position (broken line arrows).  
Model 2b is the case for no local dip configuration.  If there is a shear between the neighboring magnetic lines, magnetic reconnection can occur at that point.  In fact, Linton \& Antiochos (2005) reported that reconnection can occur even at small contact angles between the interacting magnetic field lines.  In model 2b, only upward motion will be observed.  This may account for EB9.  The 6 "top" EBs and 4 unipolar EBs on slit scan observation can be understood by both model 2a and model 2b.  
 
In model 3, there is no dips of magnetic field lines.  If there is a shear between the neighboring magnetic lines, magnetic reconnection occurs at that point.  This may account for unipolar EBs on slit scan observation.  We are not sure how velocity field is distributed in model 3.  
  
As for EB1 and EB3, inclination angles were at first uniform and gradually made dip configurations.  These two EBs appeared at the early phase of EFTs.  The dip configuration may be observed only after the EFT has developed enough, because of the spatial resolution or the formation height.  So even if we have a snapshot of magnetic inclination angle around EBs that do not show dip configuration, we can not exclude the possibility that this EB is not related to dip areas.   

Pariat et al. (2007) reported that EBs are located in regions named as bald patches, where the filed lines have geometrical form of U shape.  This result is somehow consistent with our model 1 and model 2a.  We extended their work by observing temporal evolution of magnetic configuration of an EFT.  In our data, more than half of the observed EBs can be accounted for by model 1 or model 2a.  However, there were a non-negligible number of the observed EBs that were located at the top of the $\Omega$ loop or unipolar region.  Therefore model 2b and model 3, that is, EB triggering without dip configuration are also the effective mechanisms for EBs.  

\subsection{Evidence of Parker instability}

\begin{figure}  
  \epsscale{.60}
    \plotone{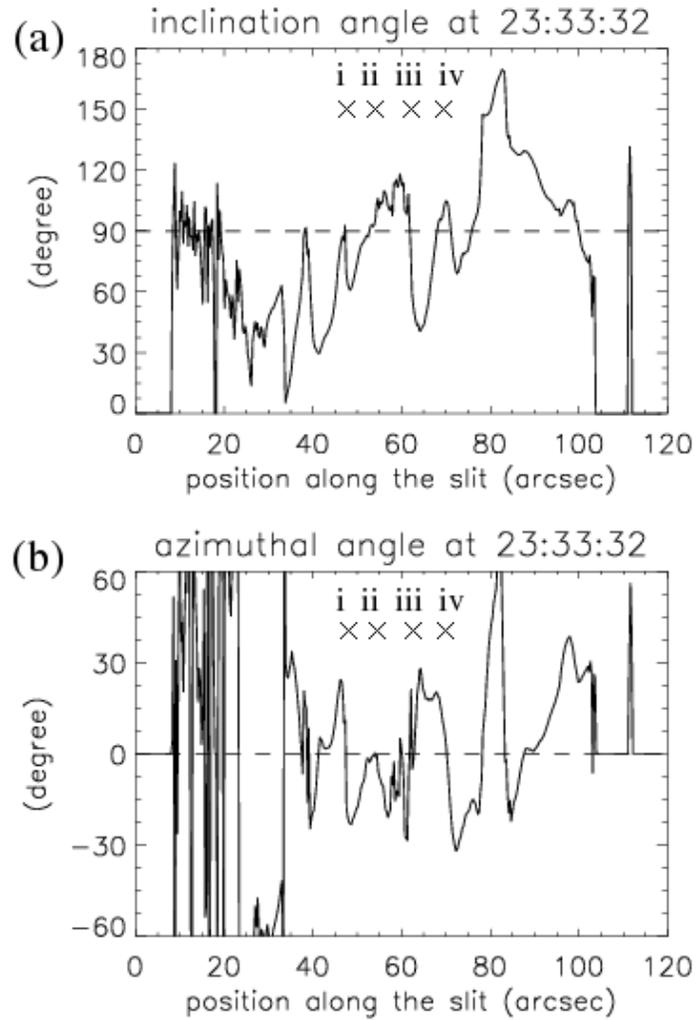}    
  \caption{(a)Magnetic inclination and (b)azimuth along slit at 23:33:32 (slit fixed).  Four cross signs are the indicators of the position of EB occurrence.  Position i (EB3, EB5, EB9).  Position ii (EB6, EB7).  Position iii (EB1, EB4, EB8).  Position iv (EB2). }\label{fig:undulation}
\end{figure}

We found an undulatory pattern whose characteristic wavelength was longer than 4000 km on both inclination and azimuth data (figure \ref{fig:undulation}).  In Pariat et al. (2004), the mechanism of an EB is magnetic reconnection of undulatory flux tubes emerging by the Parker instability \citep{key-27}.  If the undulatory magnetic fluxes are emerged by the Parker instability, magnetic field lines become undulatory in its vertical components.  Their wavelength $\lambda$ should satisfy 
\begin{equation}
\lambda > 4 \pi \textrm{H} \simeq 2000 \ \textrm{km}
\label{eq:Parker}
\end{equation}
Here H is the pressure scale height, and take value of 150-200 km at the photosphere.  This is called a resistive emergence model.  The undulatory pattern in magnetic field inclination is direct evidence for this model.  In addition, the undulatory pattern in magnetic field azimuth also prove Parker instability by the following reason.  As the magnetic fluxes that emerge by the Parker instability go up, the atmospheric pressure decreases as height and the emerging flux expands horizontally (Shibata et al. 1989).  But the footpoints anchored in photosphere are still compressed by surroundings.  When observed this EFT from the top, the horizontal components of magnetic field line shows wavy undulations with a characteristic wavelength of the Parker instability.  This is the first to report an undulation of horizontal components of magnetic field, which is supporting evidence for Parker instability.



\acknowledgments

We are grateful to all the staffs of Hida Observatory with the guidance of Domeless Solar Telescope and Vector magnetograph, and to all staffs and students of Kyoto University and Kwasan and Hida Observatories with fruitful discussions.  The authors are partially supported by a Grant-in-Aid for the 21st Century COE 'Center for Diversity and Universality in Physics' from the Ministry of Education, Culture, Sports, Science and Technology (MEXT) of Japan, and by the Grant-in-Aid for 'Creative Scientific Research The Basic Study of Space Weather Prediction' (17GS0208, Head Investigator: K. Shibata) from the Ministry of Education, Science, Sports, Technology, and Culture of Japan, and also supported by the Grant-in-Aid for the Japanese Ministry of Education, Culture, Sports, Science and Technology ( No.19540474).


\clearpage


\clearpage


\end{document}